\documentclass{article}

\usepackage{a4wide}

\usepackage{feynmp}
\usepackage{epsf}
\usepackage{color}
\usepackage{amsmath}
\usepackage{amssymb}
\usepackage{latexsym}
\usepackage{slashed}

\DeclareMathOperator{\diag}{diag}
\DeclareMathOperator{\tr}{tr}

\newcommand{\cred}[1]{#1}

\newcommand{\cl}{{^{\rm c}}}

\newcommand{\q}{{'}}

\newcommand{\true}{\text{true}}
\newcommand{\vEW}{v_\text{EW}}
\newcommand{\mKK}{m_\text{KK}}

\newcommand{\wt}{\widetilde}

\newcommand{\bW}{\boldsymbol{\mathcal W}}
\newcommand{\A}{\mathcal{A}}
\newcommand{\B}{\mathcal{B}}

\newcommand{\F}{\mathcal{F}}
\newcommand{\W}{\mathcal{W}}
\newcommand{\Z}{\mathcal{Z}}

\newcommand{\al}[1]{\begin{align}#1\end{align}}
\newcommand{\als}[1]{\begin{align*}#1\end{align*}}
\newcommand{\bp}{\begin{pmatrix}}
\newcommand{\ep}{\end{pmatrix}}
\newcommand{\bb}{\begin{bmatrix}}
\newcommand{\eb}{\end{bmatrix}}
\newcommand{\nn}{\nonumber\\}
\newcommand{\paren}[1]{\left(#1\right)}
\newcommand{\sqbr}[1]{\left[#1\right]}
\newcommand{\ab}[1]{\left|#1\right|}

\newcommand{\bs}{\boldsymbol}

\newcommand{\g}{\bs{g}}
\newcommand{\beps}{{\bs\epsilon}}

\DeclareMathOperator{\im}{Im}

\begin{document}
\begin{fmffile}{NG_letter}
% Run: ``mpost NG_letter'' in the same directory.

\title{
Unitarity in Dirichlet Higgs Model
}
\author{
	\Large
	Kenji Nishiwaki\thanks{
		E-mail: \tt nishiwaki@stu.kobe-u.ac.jp
		}
	{} and
	Kin-ya Oda\thanks{
		E-mail: \tt odakin@phys.sci.osaka-u.ac.jp
		}\medskip\\
	$^*$\it Department of Physics, Kobe University, Kobe 657-8501, Japan\smallskip\\
	$^\dagger$\it Department of Physics,  
	Osaka University,  Osaka 560-0043, Japan\smallskip\\
	}

\maketitle
\begin{abstract}\noindent
We show that a five-dimensional Universal Extra Dimension model, compactified on a line segment, is consistently formulated even when the gauge symmetry is broken solely by non-zero Dirichlet boundary conditions on a bulk Higgs field, without any quartic interaction. We find that the longitudinal $W^+W^-$ elastic scattering amplitude, under the absence of the Higgs zero mode, is unitarized by exchange of infinite towers of KK Higgs bosons. Resultant amplitude scales linearly with the scattering energy $\propto \sqrt{s}$, exhibiting five-dimensional nature. A tree-level partial-wave unitarity condition is satisfied up to $6.7\,(5.7)\,\text{TeV}$ for the KK scale $\mKK=430\,(500)\,\text{GeV}$, favored by the electroweak data within 90\% CL.
\end{abstract}
\vfill
\mbox{}\hfill CERN-PH-TH/2010-248\\
\mbox{}\hfill KOBE-TH-10-03\\
\mbox{}\hfill OU-HET-684/2010
\newpage

%%%%%%%%%%%%%%%%%%%%%%
\section{Introduction}
%%%%%%%%%%%%%%%%%%%%%%

More than four decades after the birth of the Standard Model (SM)~\cite{Weinberg:1967tq,Salam,Glashow:1970gm}, finally the CERN Large Hadron Collider (LHC) is accumulating data that will eventually reveal whether or not there exists the last missing piece of the SM, the Higgs boson, and if the Electro-Weak Symmetry Breaking (EWSB) is truly caused by the Higgs mechanism~\cite{Anderson:1963pc,Englert:1964et,Higgs:1964ia,Higgs:1964pj,Guralnik:1964eu,Higgs:1966ev,Kibble:1967sv}, namely, if or not the SM is ultimately the right description of nature at around the weak scale. The EWSB sector is the key element of the SM which eventually supplies all the masses for the elementary particles through the Yukawa couplings, but is the least experimentally confirmed part.

Even if we find a particle that looks similar to the SM Higgs boson, it is not the end of the story. It takes long way to establish whether the observed particle is really the one in the SM; see e.g.~Ref.\cite{Low:2010jp,DeRujula:2010kt}. Indeed there are many alternative EWSB mechanisms to the SM Higgs one that possess their own virtues; see e.g.~\cite{Cheng:2007bu,Grojean:2007zz} for brief overviews. Also for more reviews on Higgs/EWSB in a particular model, see e.g.\ Refs.~\cite{Serone:2005ds,Serone:2009kf} for gauge-Higgs Unification models, Ref.~\cite{Simmons:2006iw,Csaki:2005vy} for the Higgsless EWSB, Refs.~\cite{Schmaltz:2005ky,Perelstein:2005ka} for the little Higgs models, Ref.~\cite{Djouadi:2005gj} for the Minimal Supersymmetric Standard Model, and Refs.~\cite{Sannino:2008ha,Piai:2010ma} for walking technicolor models.

In Refs.~\cite{Haba:2009pb,Haba:2010xz}, it has been proposed that the EWSB can be caused without any Higgs potential if we put general non-zero Dirichlet boundary conditions on a bulk Higgs field in five dimensions, compactified on a line segment, where all the SM fields propagate in the bulk. This Dirichlet Higgs model, which is essentially the same as the Universal Extra Dimension (UED) model~\cite{Appelquist:2000nn,Appelquist:2002wb} except for the Higgs sector, predicts that there are no zero modes for the Higgs and its first Kaluza--Klein (KK) mode couples to the SM zero modes (quarks, leptons, and gauge bosons) with its coupling universally multiplied by $2\sqrt{2}/\pi\simeq 0.9$. In the first look, this Dirichlet Higgs model might appear to be equivalent to the infinitely large quartic coupling limit of the boundary-localized Higgs potential~\cite{Haba:2009uu}. However, there are no quartic coupling for the KK Higgs field in the former Dirichlet Higgs model, in contrast to the latter large boundary coupling limit that gives large quartic couplings for the KK Higgs fields. The first KK Higgs in the Dirichlet Higgs model is a ``Higgs impostor'' which has no quartic coupling and has couplings to SM sector that are always universally 10\% smaller than those in the SM.

In the Dirichlet Higgs model, the EWSB is caused by the seemingly explicit breaking at the boundaries. As we will see in Section~\ref{classical_setup}, the boundary conditions on the Higgs leave no gauge symmetry even in the bulk at the classical level. Therefore one might worry if the theory possesses a gauge symmetry at all. Furthermore, the longitudinal SM gauge bosons (zero modes) do not couple to KK gauge bosons, under the assumption that the boundary conditions respect the KK parity, i.e., when the non-zero Dirichlet boundary conditions take the same value at both boundaries.\footnote{
\cred{The zero mode gauge bosons do not couple to a pair of KK gauge bosons nor to a single KK-even gauge boson because of the accidental conservation of the KK number among the (von Neumann) gauge fields.}
}
Therefore, the KK gauge bosons do not help to unitarize the high energy behavior of the elastic scattering of longitudinal gauge bosons $W^+W^-\to W^+W^-$, unlike the Higgsless models. (Recall that there is no Higgs zero mode either.)

In this Letter, we answer above concerns. First we explain that the theory possesses a nilpotent Becchi--Rouet--Stora--Tyutin (BRST) symmetry both in five dimensions and also in a KK-expanded picture, under the non-zero Dirichlet boundary condition on the bulk Higgs field. Therefore, the Dirichlet Higgs model is fully gauge invariant as a path-integrated (or canonically quantized) quantum field theory and is unitary in the sense that there appears no unphysical degrees of freedom in external lines. 

Then we study high energy behavior of the tree-level scattering amplitude of the longitudinal SM gauge boson zero modes. We will show that the growth $\propto s$ of the elastic scattering amplitude of longitudinal $W^+W^-$ zero modes is indeed canceled by the exchanges of infinite tower of Higgs KK modes. Further, we will show that the first KK Higgs boson contributes most since the overlap of the KK wave function to zero modes decreases for higher-modes, which explains why the first KK Higgs has a coupling to all the Standard Model (SM) zero modes very close to the SM value that is multiplied by a factor $2\sqrt{2}/\pi\simeq0.9$. We also examine the partial-wave unitarity.

The organization of the paper is as follows. In Section~\ref{classical_setup}, we present the setup of our theory and show where arises a potential difficulty. Section~\ref{BG_BRST_tf} can be skipped by a reader who is not interested in formal consistency of the theory. First we explain that the background gauge transformation is viewed as a field redefinition and that the non-zero Dirichlet boundary condition can be rotated into a simpler basis. We then briefly sketch how a nilpotent BRST transformation is implemented in our model. In Section~\ref{KK_expansion_section}, we show the KK expansion of the bulk gauge, Higgs, and ghost fields. Section~\ref{scattering_section} is the main part of this Letter, where we show the high energy scattering of the longitudinal components of the zero mode gauge fields $W_L^\pm$ to exhibit the tree-level unitarity of the amplitude. In the last section, we summarize our results.

%%%%%%%%%%%%%%%%%%%%%%%%%%%%%%%%%%%%%%%%%%%%%%%%
\section{Classical setup}\label{classical_setup}
%%%%%%%%%%%%%%%%%%%%%%%%%%%%%%%%%%%%%%%%%%%%%%%%
We consider a UED model in a flat five-dimensional spacetime
\al{
ds^2
	&=	\eta_{\mu\nu}dx^\mu dx^\nu+dz^2,
}
compactified on a line segment $-L/2\leq z\leq L/2$, where indices $\mu,\nu,\dots$ run for $0,\dots,3$ and the metric signature is $\eta_{\mu\nu}=\diag\paren{-1,+1,+1,+1}$. We also let $M,N,\dots$ be five-dimensional indices running for $0,1,2,3,z$.
The gauge kinetic action is
\al{
S_g
	&=	\int d^4x\int_{-L/2}^{L/2}dz
			\sqbr{
				-{1\over2}\tr\paren{\F_{MN}\F^{MN}}
				-{1\over4}(\F^Y)_{MN}(\F^Y)^{MN}
				},
				\label{gauge_kinetic_action}
}
where
\al{
\F_{MN}
	&:=	\partial_M\W_N-\partial_N\W_M+ig[\W_M,\W_N],  \nn
(\F^Y)_{MN}
	&:=	\partial_M\B_N-\partial_N\B_M,
}
$\B_M$ is the $U(1)_Y$ gauge field and $\W_M:=\W^a_MT^a$, with $a,b,\dots$ running for $SU(2)_W$ adjoint indices $1,2,3$ whose summation is being understood unless otherwise stated, and $[T^a,T^b]=i\epsilon^{abc}T^c$; We have normalized to $\tr\paren{T^aT^b}=1/2$, as usual. We also write collectively
\al{
\bW_M
	&:=	\sum_A\W_M^AT^A,	&
\g\bW_M
	&:=	\sum_Ag_A\W_M^AT^A,
}
where $A$ run for $1,2,3,Y$ with $g_1=g_2=g_3:=g$, and correspondingly $\W^Y_M:=\B_M$ and $T^Y:=Y$.

The Higgs action is
\al{
S_\Phi
	&=	\int d^4x\int_{-L/2}^{L/2}dz
	\sqbr{
	-\paren{D_M\Phi}^\dagger D^M\Phi
	-V(\Phi)
	},
		\label{Higgs_action}
}
where
\al{
D_M\Phi
	&:=	\paren{\partial_M+i\g\bW_M}\Phi.
}
On $\Phi$, $Y=1/2$ and $T^a=\sigma^a/2$ with $\sigma^a$ being Pauli matrices.
In this Letter we set $V(\Phi)=0$ since we are interested in the theoretical consistency of putting the general non-zero Dirichlet boundary condition on the Higgs field~\cite{Haba:2009pb,Haba:2010xz}. Essential features such as BRST invariance and unitarization of longitudinal gauge boson scattering are not altered by inclusion of $V(\Phi)$.

On all the gauge fields, we put the standard von Neumann and Dirichlet boundary conditions on $\A_\mu(x,z)$ and $\A_z(x,z)$, respectively, at both ends of the line segment:
\al{
\partial_z\W^A_\mu(x,\pm L/2)
	&=	0,	&
\W^A_z(x,\pm L/2)
	&=	0.	\label{A_bc}
}
On the Higgs field $\Phi(x,z)$, we impose the most general non-zero Dirichlet boundary condition~\cite{Haba:2009pb,Haba:2010xz}:
\al{
\Phi(x,\pm L/2)
	&=	\bb
		\phi_D^1\\
		\phi_D^2
		\eb
	=:	\Phi_D,
	\label{Higgs_bc}
}
where $\phi_D^1$ and $\phi_D^2$ are arbitrary complex constants and we have assumed that the KK parity $z\to-z$ is preserved by the boundary conditions, $\Phi(x,L/2)=\Phi(x,-L/2)$, for simplicity.
Note that, without loss of generality, we can perform a field redefinition to rotate the boundary condition to become
\al{
\Phi_D
	&\to
		\Phi_D^\text{new}
			=	\bb
				0\\
				v/\sqrt{2}
				\eb,
				\label{bc_final_form}
}
where $v$ is a real parameter, but we leave it general as in Eq.~\eqref{Higgs_bc} for the moment to see below how the background gauge invariance is implemented in the Dirichlet Higgs model.

For our purpose, is it most convenient to employ the background field method, see e.g.\ Ref.~\cite{Denner:1994xt}, in which we separate a field into a classical background and a quantum fluctuation around it:
\al{
\Phi
	&=	\Phi^\cl +\Phi^\q, &
\bW_M
	&=	\bW_M^\cl +\bW_M^\q.
}
Throughout this paper,  $'$ on a field does not denote derivative. The classical equation of motion for $\Phi^\cl (x,z)$ is given by the variation in the bulk as
\al{
\paren{\Box+\partial_z^2}\Phi^\cl (x,z)=0,
\label{classical_eom}
}
where $\Box:=\partial_\mu\partial^\mu$.
An obvious classical solution to the e.o.m.~\eqref{classical_eom} under the boundary condition~\eqref{Higgs_bc} is the constant one
\al{
\Phi^\cl (x,z)=\Phi_D.
	\label{Higgs_vev}
}
Around this vacuum expectation value (vev), the Higgs field is now expanded as
\al{
\Phi(x,z)
	&=	\Phi_D+\Phi^\q(x,z),
}
Let us emphasize that the non-zero Dirichlet boundary condition~\eqref{Higgs_bc} implies that the boundary condition for the quantum fluctuation reduces to the ordinary vanishing Dirichlet condition
\al{
\Phi^\q(x,\pm L/2)
	&=	0.
		\label{Dirhichlet_on_fluctuation}
}

We note that, at classical level (omitting $\cl$), a gauge transformation in five dimensions,
\al{
\Phi(x,z)
	&\to e^{i\g\bs\theta(x,z)}\Phi(x,z),\nn
i\g\bW_M(x,z)
	&\to	e^{i\g\bs\theta(x,z)}\paren{\partial_M+i\g\bW_M(x,z)}e^{-i\g\bs\theta(x,z)},
}
where $\g\bs\theta(x,z):=\sum_Ag_A\theta^A(x,z)T^A$, does not change the boundary conditions on the gauge fields~\eqref{A_bc} when and only when all the gauge parameters satisfy the von Neumann condition:
\al{
\partial_z\theta^A(x,\pm L/2)
	&=	0.
		\label{bc_on_gauge_parameter}
}
However, for a general non-zero Dirichlet boundary condition on Higgs~\eqref{Higgs_bc}, 
%it appears that the only remaining gauge transformation is the (vanishing) Dirichlet one $\theta^A(x,\pm L/2)=0$, which is apparently in contradiction to Eq.~\eqref{bc_on_gauge_parameter}. 
it appears that the broken gauge parameter for $SU(2)_W/U(1)_\text{EM}$ must also obey the (vanishing) Dirichlet condition $\theta^A(x,\pm L/2)=0$, which, with Eq. (16), shows that $\theta^A(x,z)=0$ everywhere.
It looks as if the symmetry breaking by the conditions~\eqref{A_bc} and \eqref{Higgs_bc} were an explicit breaking and there remained no $SU(2)_W$ symmetry even in the bulk of five-dimensional space.
By this classical argument, the theory looks pathetic.
How can we overcome this difficulty?

The key observation is that the Dirichlet boundary condition on the Higgs field fluctuation~\eqref{Dirhichlet_on_fluctuation} remains to be Dirichlet when multiplied by a function $\theta(x,z)$ with von Neumann condition~\eqref{bc_on_gauge_parameter}, that is, the condition~\eqref{Dirhichlet_on_fluctuation} is preserved by the von Neumann transformation $\theta(x,z)$.
We will see how this observation is implemented as the nilpotent BRST transformation in the following.

%%%%%%%%%%%%%%%%%%%%%%%%%%%%%%%%%%%%%%%%%%%%%%%%%%%%%%%%%%%%%%%%%%
\section{Background and BRST transformations}\label{BG_BRST_tf}
%%%%%%%%%%%%%%%%%%%%%%%%%%%%%%%%%%%%%%%%%%%%%%%%%%%%%%%%%%%%%%%%%%
In this section, we briefly outline how the theory is consistently defined. A reader who is not interested in formal consistency may skip the entire section.
%%%%%%%%%%%
\subsection{Background-field $R_\xi$ gauge fixing}
%%%%%%%%%%%
We employ the following gauge fixing, the background-field $R_\xi$ gauge:
\al{
S_\xi
	&=	\int d^4x\int_{-L/2}^{L/2}dz
		\sqbr{-{1\over2\xi}f^Af^A
			},
		\label{gauge_fixing}
}
with $A$ running for $1,2,3,Y$ and the gauge fixing function being given by
\al{
f^A
	&:=	D_\mu^\cl\W^{\q A\mu}+\xi D_z^\cl\W^{\q Az}
		+ig_A\xi\paren{(\Phi^\q)^\dagger T^A\Phi^\cl-(\Phi^\cl)^\dagger T^A\Phi^\q},
		\label{gauge_fixing_function}
}
where $g_1=g_2=g_3=:g$, $T^Y:=Y$, $\W^Y_M:=\B_M$, $\xi$ is a dimensionless positive constant, and we define the background covariant derivative on an arbitrary $SU(2)_W$ adjoint field $\Phi_\text{ad}$ as $D^\cl _M\Phi_\text{ad}:=\partial_M\Phi_\text{ad}+ig[\W^\cl _M,\Phi_\text{ad}]$. Note that $D^\cl_M\B_N=\partial_M\B_N$.

The true gauge transformation that is fixed by the gauge choice~\eqref{gauge_fixing_function} is, in its infinitesimal form,
\al{
\Delta^\true_\beps \W_M^{\q A} 
	&=	-D^\cl_M\epsilon^A+i[\g\beps,\bW^\q_M]^A, &
\Delta^\true_\beps \W_M^{\cl A} 
	&=	0,	\nn
\Delta^\true_\beps \Phi^\q
	&=	i\g\beps\,(\Phi^\cl +\Phi^\q),	&
\Delta^\true_\beps \Phi^\cl 
	&=	0,	
		\label{true_gauge_transformation}
}
with $\g\beps:=\sum_Ag_A\epsilon^AT^A$, from which the ghost Lagrangian can be read off as
\al{
\mathcal{L}_\omega
	&=	-\bar\omega^A\Delta^\true_{\bs\omega}f^A\nn
	&=	-\bar\omega^AD^{\cl\mu}\paren{
			-D^\cl_\mu\omega^A
			+i[\g\bs\omega,\bW^\q_\mu]^A
			}
		-\xi\bar\omega^AD_z^\cl\paren{
			-D^\cl_z\epsilon^A
			+i[\g\bs\omega,\bW^\q_z]^A
			}\nn
	&\quad
		-\xi\paren{
			-(\Phi^\cl+\Phi^\q)^\dagger
				\paren{\g\bs\omega}
				\paren{\g\bar{\bs\omega}}
				\Phi^\cl
			+(\Phi^\cl)^\dagger
				\paren{\g\bar{\bs\omega}}
				\paren{\g\bs\omega}
				(\Phi^\cl+\Phi^\q)
			},\label{ghost_L}
}
where $\g\bs\omega:=\sum_Ag_A\omega^AT^A$ and $\g\bar{\bs\omega}:=\sum_Ag_A
\bar\omega^AT^A$.

The background gauge transformation is given, with $\bs\omega:=\sum_A\omega^AT^A$, by
\al{
\delta \W_M^{\q A} 
	&=	i[\g\beps,\bW^\q _M]^A,	&
\delta \W_M^{\cl A} 
	&=	-D^\cl _M\epsilon^A,	\nn
\delta \Phi^\q
	&=	i\g\beps\Phi^\q,	&
\delta \Phi^\cl 
	&=	i\g\beps\Phi^\cl,\nn
\delta\omega^{\q A}
	&=	i[\g\beps,\bs\omega^\q]^A,	&
\delta\omega^{\cl A}
	&=	i[\g\beps,\bs\omega^\cl]^A, \nn
\delta\bar\omega^{\q A}
	&=	i[\g\beps,\bs{\bar\omega}^\q]^A,&
\delta\bar\omega^{\cl A}
	&=	i[\g\beps,\bs{\bar\omega}^\cl]^A,
	\label{background_gauge_transformation}
}
which transforms (anti-)ghost and the quantum fluctuation $\W_M'$ as adjoint and leaves the ghost Lagrangian~\eqref{ghost_L} manifestly invariant.
Noting that the background transformation~\eqref{background_gauge_transformation} varies the gauge-fixing function as adjoint: 
\al{
\delta f^A
	&=	i[\g\beps,\bs f]^A,
}
we find that the total action, i.e.\ the gauge fixing action~\eqref{gauge_fixing} as well as the original gauge~\eqref{gauge_kinetic_action} and Higgs~\eqref{Higgs_action} actions are invariant under the background gauge transformation~\eqref{background_gauge_transformation}.

Note that the rotated field by the transformation~\eqref{background_gauge_transformation} satisfies the following boundary condition:
\al{
\Phi^\q(x,\pm L/2)^\text{new}
	&=	0,\\
\Phi^\cl (x,\pm L/2)^\text{new}
	&=	e^{i\g\beps(x,\pm L/2)}\Phi^\cl (x,\pm L/2)
	=	e^{i\g\beps(x,\pm L/2)}\Phi_D,
}
that is, the quantum fluctuation does not change its boundary condition (b.c.) by the background transformation though the vev does change its b.c.\ into
\al{
\Phi_D^\text{new}(x,\pm L/2)
	&=	e^{i\g\beps(x,\pm L/2)}\Phi_D.
}
This is natural since the background transformation~\eqref{background_gauge_transformation} rotates the vevs $\Phi^\cl $ and $\A_M^\cl $ and hence should be regarded as a field redefinition, unlike the true gauge transformation~\eqref{true_gauge_transformation}. The field redefinition certainly must change the b.c.

When we consider a background transformation (namely field redefinition) that respects the KK parity $\epsilon^A(x,L/2)=\epsilon^A(x,-L/2)$, the rotated boundary conditions remain to respect it too $\Phi_D^\text{new}(x,L/2)=\Phi_D^\text{new}(x,-L/2)$. In particular, by a global background transformation
\al{
\Phi^\q(x,z)
	&\to	e^{i\g\beps}\Phi^\q(x,z),	&
\Phi^\cl (x,z)
	&\to	e^{i\g\beps}\Phi^\cl(x,z), \nn
\bW^\cl_M(x,z)
	&\to	e^{i\g\beps}\bW^\cl_M(x,z)e^{-i\g\beps},&
\bW^\q_M(x,z)
	&\to	e^{i\g\beps}\bW^\q_M(x,z)e^{-i\g\beps},
}
the boundary condition for Higgs can always be rotated to the form~\eqref{bc_final_form}.

%%%%%%%%%%%
\subsection{BRST invariance}
%%%%%%%%%%%
The bulk BRST transformation can be introduced quite the same way as in the four-dimensional (4D) gauge theory. On physical degrees of freedom, it is defined as a true gauge transformation with its gauge parameter being replaced by the ghost field:
\al{
s\W_M^{\q A} 
	&=	-\partial_M\omega^A+i[\g\bs\omega,\bW^\q_M]^A, &
s\W_M^{\cl A} 
	&=	0,	\nn
s\Phi^\q
	&=	i\g\bs\omega\,(\Phi^\cl +\Phi^\q),	&
s\Phi^\cl 
	&=	0.
		\label{BRST_transformation}
}
On unphysical fields, the BRST transformation reads
\al{
s\omega^A
	&=	{i\over2}[\g\bs\omega,\bs\omega]^A,	&
s\bar\omega^A
	&=	h^A,	&
sh^A
	&=	0,
}
where we take $\omega^{\cl A}=\bar\omega^{\cl A}=h^{\cl A}=0$ and drop $\q$ from the quantum fluctuations. We see that the action, including the gauge fixing and ghost terms, is invariant under the BRST transformation~\eqref{BRST_transformation}.

The only non-triviality here is the appearance of $\Phi^\cl$ in the transformation of $\Phi^\q$ but it is still straightforward to show the nilpotency of the BRST transformation on $\Phi^\q$. One might worry that the flat configuration $\Phi^\cl$ is mixed with the Dirichlet field $\Phi^\q$ after the transformation. To answer it, we can KK-expand the transformation~\eqref{BRST_transformation} and define it on the expanded fields. More detailed explanation will be shown in a separate publication~\cite{NishiwakiOda}.\footnote{
In~\cite{Ohl:2003dp}, a higher-dimensional BRST symmetry is considered for orbifold gauge theories.
In~\cite{Abe:2004wv}, an orbifold GUT is studied with infinite number of 4D gauge-fixing terms, where a BRST symmetry is proposed including the corresponding infinite number of 4D ghosts, with its nilpotency being untouched.
}

%%%%%%%%%%%%%%%%%%%%%%%%%%%%%%%%%
\section{KK expansions}\label{KK_expansion_section}
%%%%%%%%%%%%%%%%%%%%%%%%%%%%%%%%%

From now on, we choose the basis in which the b.c.\ becomes~\eqref{bc_final_form}, which leads to the vev
\al{
\Phi^\cl (x,z)
	&=	\bb
		0\\
		{v/\sqrt{2}}
		\eb,
		\label{vev_in_particular_basis}
}
where $v:=\sqrt{2}\paren{|\varphi^1_D|^2+|\varphi^2_D|^2}^{1/2}$ in terms of the original most general boundary condition~\eqref{Higgs_bc}.
Let us rewrite the Higgs fluctuation as:
\al{
\Phi^\q(x,z)
	&=	\bb
		\chi^+(x,z)\\
		{\varphi(x,z)+i\chi(x,z)\over\sqrt{2}}
		\eb,
}
where we omit $'$ from fluctuations $\varphi$, $\chi^+$, and $\chi$. The boundary condition is now
\al{
\chi^\pm(x,\pm L/2)
	=	\varphi(x,\pm L/2)
	=	\chi(x,\pm L/2)
	=	0.
	\label{bc_for_phichi_ADHM}
}
On physical ground, we put $\W_M^{\cl A}=\omega^{\cl A}=\bar\omega^{\cl A}=0$ hereafter (and drop $'$ from the quantum fluctuations unless otherwise stated).\footnote{
Since we are putting the (vanishing) Dirichlet boundary condition on $\W^A_z$, we do not have $\W^{\cl A}_z$ nor the Wilson line along the extra dimension.
}
Then gauge fields in the mass eigenbasis are, as usual,
\al{
\W^\pm_M
	&:=	{1\over\sqrt{2}}\paren{\W^1_M\mp i\W^2_M},	&
\bb
\Z_M\\
\A_M
\eb
	&:=	\bb
		\cos\theta_W	&	-\sin\theta_W\\
		\sin\theta_W	&	\cos\theta_W
		\eb
		\bb
		\W^3_M\\
		\B_M
		\eb,
}
where $\sin\theta_W:=g_Y/\sqrt{g^2+g_Y^2}$.

After some manipulations, all the von Neumann and Dirichlet fields $\Psi^N$ and $\Psi^D$, respectively, are KK-expanded as~\cite{NishiwakiOda}\footnote{\label{zero_mode_normalization}
In this notation, a zero mode becomes canonically normalized in terms of a redefined field $\psi^N_n(x)$, where the KK modes are normalized by $\psi^N_0(x):=\Psi^N_0(x)/\sqrt{2}$ for $n=0$ and by $\psi^N_n(x):=\Psi^N_n(x)$ for $n\neq0$. We note that we are defining the negative KK modes by $\Psi^N_{-n}(x)=\Psi^N_n(x)$ and $\Psi^D_{-n}(x)=-\Psi^D_n(x)$, which is consistent with the choice of the normalization $C_{-n}(z)=C_n(z)$ and $S_{-n}(z)=-S_n(z)$.
}
\al{
\Psi^N(x,z)
	&=	\sum_{n=-\infty}^\infty C_n(z)\Psi^N_n(x), &
\Psi^D(x,z)
	&=	\sum_{n=-\infty}^\infty S_n(z)\Psi^D_n(x),
}
where
\al{
C_n(z)
	&:=	{1\over\sqrt{2L}}\cos\!\sqbr{
			{n\pi\over L}\paren{
				z+{L\over2}
				}
			}
	=	{1\over\sqrt{2L}}\times
		\begin{cases}
		(-1)^{n\over2}\cos{n\pi z\over L}
			&	\text{for $n$: even,}\\
		(-1)^{n+1\over2}\sin{n\pi z\over L}
			&	\text{for $n$: odd,}
		\end{cases}\nn
S_n(z)
	&:=	{1\over\sqrt{2L}}\sin\!\sqbr{
			{n\pi\over L}\paren{
				z+{L\over2}
				}
			}
	=		{1\over\sqrt{2L}}\times
		\begin{cases}
		(-1)^{n\over2}\sin{n\pi z\over L} & \text{for $n$: even,}\\
		(-1)^{n-1\over2}\cos{n\pi z\over L} & \text{for $n$: odd.}
		\end{cases}
		\label{KK_modes_defined}
}
Concretely, the von Neumann boundary condition is satisfied by all the gauge fields $\W^\pm_\mu$, $\Z_\mu$, $\A_\mu$ and ghost fields (as well as all the quarks and leptons), whereas the (vanishing) Dirichlet boundary condition is satisfied by all the Higgs fluctuations $\varphi^\pm$, $\varphi$, $\chi$ and all the vector-scalars $\W^\pm_z$, $\Z_z$, $\A_z$. A crucial point is that fields with von Neumann and non-zero Dirichlet boundary conditions are not necessarily orthogonal to each other though Dirichlet function is orthogonal to Dirichlet ones and vice versa, as a line segment does not have periodicity.
This feature becomes important in the next section.

% In the unitary gauge $\xi\to\infty$, we find that all the ghost fields as well as the NG-bosons $\wt\chi^\pm:=\chi^\pm+\partial_z\W^\pm_z/m_W$ and $\wt\chi:=\chi+\partial_z\Z_z/m_Z$ become infinitely heavy and decouple, where $m_W:=gv/2$ and $m_Z:=\sqrt{g^2+g_Y^2}v/2$. Furthermore, the other linear combinations $\wt\W^\pm_z:=\W^\pm_z+\partial_z\chi^\pm/m_W$ and $\wt\Z_z:=\Z_z+\partial_z\chi/m_Z$ are imposed the boundary conditions $\wt\W^\pm_z(x,\pm L/2)=\partial_z\wt\W^\pm_z(x,\pm L/2)=0$ and $\wt\Z_z(x,\pm L/2)=\partial_z\wt\Z_z(x,\pm L/2)=0$ in the limit $\xi\to\infty$ and therefore decouple too, since $\wt\W = \wt\Z  = 0$ is derived anywhere in the bulk by the above boundary conditions.
% To summarize, the physical degrees of freedom are $\W_\mu^\pm$, $\Z_\mu$, $\A_\mu$, and $\varphi$ in the gauge-Higgs sector. Finally, 
We find that the KK masses for physical degrees of freedom are~\cite{NishiwakiOda}
\al{
\mu^2_W
	&=	m_W^2+{n^2\over R^2}	&
	&(n\geq0),\nn
\mu^2_Z
	&=	m_Z^2+{n^2\over R^2}	&
	&(n\geq0),\nn
\mu^2_{\varphi}
	&=	{n^2\over R^2}	&
	&(n\geq1).
}
where $L=:\pi R$. Note that $S_0(z)=0$ and there are no zero mode for a Dirichlet field. In particular, it is important that there is no zero mode for the physical Higgs field $\varphi$ because it obeys the Dirichlet boundary condition~\cite{Haba:2009pb,Haba:2010xz}. Below we will see how the elastic scattering of longitudinal $W^+W^-$ zero modes is unitarized in high energies in our model where we do not have a Higgs zero mode.

%%%%%%%%%%%%%%%%%%%%%%%%%%%%%%%%%%%%%%%%%%%%%%
\section{Unitarity in elastic scattering}\label{scattering_section}
%%%%%%%%%%%%%%%%%%%%%%%%%%%%%%%%%%%%%%%%%%%%%%

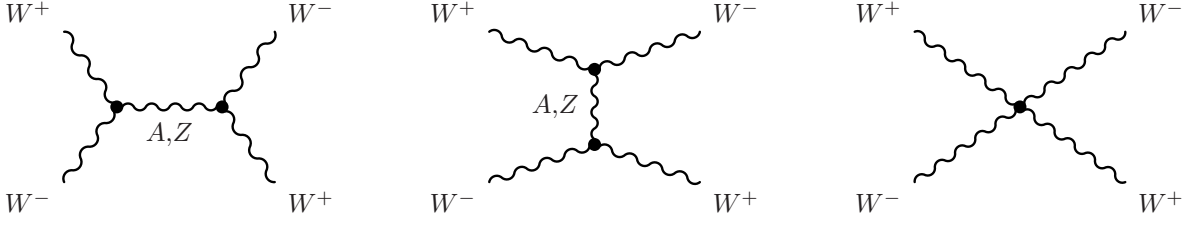
\begin{figure}
\begin{fmfgraph*}(35,20)
\fmfleft{i1,i2}
\fmfright{o1,o2}
\fmflabel{$W^-$}{i1}
\fmflabel{$W^+$}{i2}
\fmflabel{$W^+$}{o1}
\fmflabel{$W^-$}{o2}
\fmflabel{}{v1}
\fmflabel{}{v2}
\fmf{photon}{i1,v1,i2}
\fmf{photon}{o1,v2,o2}
\fmf{photon,label=$A$,,$Z$}{v1,v2}
\fmfdot{v1,v2}
\end{fmfgraph*}
\hfill
\begin{fmfgraph*}(35,20)
\fmfleft{i1,i2}
\fmfright{o1,o2}
\fmflabel{$W^-$}{i1}
\fmflabel{$W^+$}{i2}
\fmflabel{$W^+$}{o1}
\fmflabel{$W^-$}{o2}
\fmflabel{}{v1}
\fmflabel{}{v2}
\fmf{photon}{i1,v1}
\fmf{photon}{i2,v2}
\fmf{photon}{v1,o1}
\fmf{photon}{v2,o2}
\fmf{photon,label=$A$,,$Z$}{v1,v2}
%\fmffreeze\fmfdraw
%\fmf{photon}{v1,o2}
%\fmf{photon}{v2,o1}
\fmfdot{v1,v2}
\end{fmfgraph*}
\hfill
\begin{fmfgraph*}(35,20)
\fmfleft{i1,i2}
\fmfright{o1,o2}
\fmflabel{$W^-$}{i1}
\fmflabel{$W^+$}{i2}
\fmflabel{$W^+$}{o1}
\fmflabel{$W^-$}{o2}
\fmflabel{}{v}
\fmf{photon}{i1,v}
\fmf{photon}{i2,v}
\fmf{photon}{o1,v}
\fmf{photon}{o2,v}
\fmfdot{v}
\end{fmfgraph*}
\vspace{0.5cm}
\caption{SM gauge interactions involving only zero modes, where charges are written as all incoming.}
\label{SM_gauge_fig}
\end{figure}

Let us consider the elastic scattering of longitudinal modes $W^+_LW^-_L\to W^+_LW^-_L$. In the absence of the Higgs zero mode, the SM contributions to the gauge boson scattering amplitude, shown in Fig.~\ref{SM_gauge_fig}, grows with energy as~\cite{Lee:1977eg}
\al{
\mathcal{M}^\text{SM gauge only}_{W_L^+W_L^-\to W_L^+W_L^-}
	&=	{s\paren{1+\cos\theta}\over 2\vEW^2}+\mathcal{O}(s^0),
		\label{SM_gauge_only}
}
where $\vEW\simeq 246\,\text{GeV}$ is the electroweak scale, $\theta$ is the scattering angle in CM frame and $s$ is the Mandelstam variable. Note that in our notation, $\vEW=v\sqrt{L}$.

In the Higgsless model, KK modes of the gauge fields served to unitarize this high energy behavior. In our model with the KK parity respecting boundary condition, no KK mode of gauge/vector-scalar fields can couple to the external zero mode $W^\pm$~\cite{NishiwakiOda}. Then what can unitarize the $W^+_LW^-_L$ scattering in our model, where there are no zero mode Higgs? Hereafter, we show that infinite tower of the Higgs KK modes $\varphi_n(x)$ do unitarize the scattering of longitudinal modes.

%%%%%%%%%%%
\subsection{KK Higgs exchange amplitude}
%%%%%%%%%%%

In our model, the KK parity of the physical Higgs field becomes flipped from that of a von Neumann field. Furthermore, as a result of non-orthogonality of Dirichlet and von Neumann fields, the odd KK Higgs field can have a coupling to the longitudinal $W^\pm$ zero mode:
% Feynman rule: （ラグランジアンを$i$倍する。）
\vspace{0.5cm}
\al{
\parbox{40mm}{
	\begin{fmfgraph*}(20,15)%\fmfkeep{AAphi_vertex}
		\fmfleft{i1,i2}\fmfright{o1}
		\fmflabel{$W^{+\mu}$}{i2}
		\fmflabel{$W^{-\nu}$}{i1}
		\fmflabel{$\varphi_n$}{o1}
		\fmf{photon}{i1,v,i2}
		\fmf{dashes}{o1,v}
		\fmfdot{v}
	\end{fmfgraph*}
	}
	&=	{-2\sqrt{2}i\over n\pi}g_4m_W\eta_{\mu\nu},\\
\nonumber
}
where $n>0$ is a positive odd integer, $g_4:=g/\sqrt{L}$ is the four-dimensional $SU(2)_W$ gauge coupling, and $W^{\pm\mu}(x):=\W^{\pm\mu}_0(x)/\sqrt{2}$ is the canonically normalized zero mode; see footnote~\ref{zero_mode_normalization}. We note that the coupling of the $n$th KK Higgs mode is multiplied by the factor $2\sqrt{2}/n\pi\simeq 0.9/n$. In particular, the first KK Higgs mode coupling to all the zero mode SM fermions and gauge bosons are multiplied by this factor $2\sqrt{2}/\pi\simeq 0.9$. We will discuss below why this first KK Higgs behaves almost like the SM Higgs, though it has no quartic coupling.

\begin{figure}
\begin{center}
\vspace{0.5cm}
\hfill
\begin{fmfgraph*}(35,20)
\fmfleft{i1,i2}
\fmfright{o1,o2}
\fmflabel{$W^-$}{i1}
\fmflabel{$W^+$}{i2}
\fmflabel{$W^+$}{o1}
\fmflabel{$W^-$}{o2}
\fmflabel{}{v1}
\fmflabel{}{v2}
\fmf{photon}{i1,v1,i2}
\fmf{photon}{o1,v2,o2}
\fmf{dashes,label=$\varphi_n$}{v1,v2}
\fmfdot{v1,v2}
\end{fmfgraph*}
\hfill
\begin{fmfgraph*}(35,20)
\fmfleft{i1,i2}
\fmfright{o1,o2}
\fmflabel{$W^-$}{i1}
\fmflabel{$W^+$}{i2}
\fmflabel{$W^+$}{o1}
\fmflabel{$W^-$}{o2}
\fmflabel{}{v1}
\fmflabel{}{v2}
\fmf{photon}{i1,v1}
\fmf{photon}{i2,v2}
\fmf{photon}{v1,o1}
\fmf{photon}{v2,o2}
\fmf{dashes,label=$\varphi_n$}{v1,v2}
%\fmffreeze\fmfdraw
%\fmf{photon}{v1,o2}
%\fmf{photon}{v2,o1}
\fmfdot{v1,v2}
\end{fmfgraph*}
\hfill
\vspace{0.5cm}
\caption{$s$ and $t$-channel KK Higgs exchange diagrams, where charge convention is given as in Fig.~\ref{SM_gauge_fig}. $n>0$ is odd.}
\label{KK_Higgs_fig}
\end{center}
\end{figure}
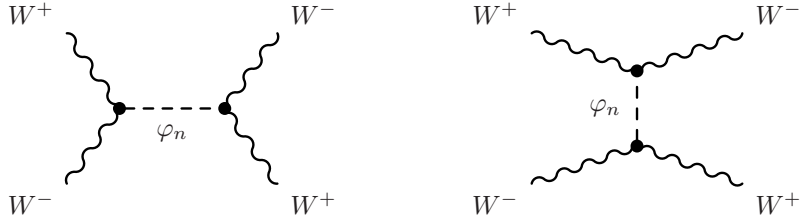

The $s$ and $t$-channel Higgs-exchange diagrams are shown in Fig.~\ref{KK_Higgs_fig}.
In the Feynman-'t Hooft gauge $\xi=1$, we can check that these are the only additional diagrams and get %\footnote{
%メトリックのconventionはワインバーグ、それ以外$2\pi$ factorsとか外線factorsとかのconventionはペスキンに従う。
%}
\al{
\mathcal{M}_{W^+_LW^-_L\to W^+_LW^-_L}^\text{KK Higgs exchange}
	&=	-\sum_\text{$n>0$, odd}
			{8g_4^2m_W^2\over n^2\pi^2}
			\sqbr{
				{\paren{1-{s\over2m_W^2}}^2
					\over s-\paren{n\over R}^2}
%				+{\paren{1-\paren{2+{2t\over s-4m_W^2}}{s\over4m_W^2}}^2
				+{\paren{1+{2t\over s-4m_W^2}{s\over4m_W^2}}^2
					\over t-\paren{n\over R}^2}
				},
}
where $t=-\paren{s-4m_W^2}\paren{1-\cos\theta}/2$.
When we take the hard scattering limit with large $s$ and fixed scattering angle $\theta$ for each contribution from the $n$th KK Higgs mode,
\al{
\mathcal{M}_{W^+_LW^-_L\to W^+_LW^-_L}^\text{KK Higgs exchange}
	&=	-\sum_\text{$n>0$, odd}\paren{2\sqrt{2}\over n\pi}^2{s\paren{1+\cos\theta}\over2\vEW^2}+\mathcal{O}(s^0).
		\label{amplitude_before_summation_in_limit}
}
As stated above, the first KK Higgs almost ($\simeq 81\%$) cancels the SM gauge contribution~\eqref{SM_gauge_only} because the higher KK modes have smaller overlapping with the von Neumann zero mode and the first one contributes most. This is why the first KK Higgs behaves almost like the SM Higgs with all its coupling to SM zero modes multiplied by ${2\sqrt{2}\over\pi}\simeq0.9$. It almost unitarizes the $WW$ scattering, hence it is almost a Higgs. Finally by performing the summation, we get
\al{
\mathcal{M}_{W^+_LW^-_L\to W^+_LW^-_L}^\text{KK Higgs exchange}
	&=	-{s\paren{1+\cos\theta}\over2\vEW^2}+\mathcal{O}(s^0),
		\label{summed_amplitude_in_limit}
}
which exactly cancels and unitarizes the SM gauge contribution~\eqref{SM_gauge_only}.

In general, an elastic scattering amplitude of massive gauge bosons is expanded as
\al{
\mathcal M
	&=	{s^2\over{\vEW^4}}\mathcal{M}^{(4)}
		+{s\over{\vEW^2}}\mathcal{M}^{(2)}
		+\mathcal{M}^{(0)}
		+\mathcal{O}\paren{{\vEW^2}\over s}.
		\label{general_expansion}
}
If the non-zero Dirichlet b.c.\ were not put on a Higgs, $\vEW$ would be replaced by $\mKK:=1/R$ generally in Eq.~\eqref{general_expansion}. In such an expansion, cancelation of $\mathcal{O}(s^2)$ and $\mathcal{O}(s)$ terms has been shown for gauge theories on $S^1/Z_2$~\cite{SekharChivukula:2001hz}, for an electroweak $SU(3)_W$ model\footnote{
The bulk $SU(3)_W$ is broken down to $SU(2)_W\times U(1)_Y=:G_\text{SM}$, and the high energy scattering unitarity of KK gauge bosons $W^{(1/2)}$, which belong to the broken non-SM sector $SU(3)_W/G_\text{SM}$, is verified under the assumption that $W^{(1/2)}$ had the same interaction to $\gamma,Z$ as that of the SM $W^\pm$ living in the unbroken $G_\text{SM}$~\cite{Abe:2003vg}.
}
and an $SU(5)$ GUT model on the orbifold $S^1/Z_2$~\cite{Abe:2003vg}, and for Higgsless models on $S^1/Z_2$~\cite{Chivukula:2003kq} and on a line segment~\cite{Csaki:2003dt}, all of which are equivalent to taking the limit~\eqref{amplitude_before_summation_in_limit} before summation.
%, as there are only few numbers of modes that actually contribute. 
In our case, we have seen that the terms of $\mathcal{O}(s^2)$ cancels within SM gauge amplitudes, while the sum over the terms of $\mathcal{O}(s)$ from the SM gauge sector~\eqref{SM_gauge_only} is canceled by the infinite sum over all the odd-$n$ KK Higgs modes~\eqref{summed_amplitude_in_limit}. Actually, we can go one step further from the analysis of Refs.~\cite{SekharChivukula:2001hz,Abe:2003vg,Chivukula:2003kq,Csaki:2003dt}. Let us see it below.

One might still worry that the high energy limit $s\to\infty$ is taken before the infinite summation. We can indeed exactly perform the infinite sum before taking the limit, so as not to spoil five-dimensional symmetries:
\al{
\mathcal{M}_{W^+_LW^-_L\to W^+_LW^-_L}^\text{KK Higgs exchange}
	&=	-{s\over\vEW^2}\paren{1-{2m_W^2\over s}}^2\sqbr{
			1
			-{2\over\pi R\sqrt{s}}\tan{\pi R\sqrt{s}\over 2}
			}
			\nn
	&\quad
		+{\ab{t}\over\vEW^2}\paren{{1\over1-{4m_W^2\over s}}-{2m_W^2\over \ab{t}}}^2
			\sqbr{
				1
				-{2\over\pi R\sqrt{\ab{t}}}\tanh{\pi R\sqrt{\ab{t}}\over2}
				},
			\label{exact_sum}
}
where $-s+4m_W^2\leq t\leq0$. In the hard scattering limit $s\to \infty$ with fixed scattering angle $\theta$, the hyperbolic tangent goes to unity exponentially:
\al{
\tanh{{\pi R\sqrt{\ab{t}}\over 2}}\to 1.
}
How about the tangent: $\tan{\pi R\sqrt{s}\over 2}$? We see that there appear poles at $\sqrt{s}=n/R=:m_n$ ($n=1,3,\dots$), which are nothing but the remnant of the $s$-channel $\varphi_n$ resonance production. In string theory, we know how to treat this kind of infinite number of poles. If we take the higher loop corrections into account, these poles on the real axis of complex $s$ plane will be shifted to
\al{
{1\over s-m_n^2}
	\to	{1\over s-m_n^2+im_n\Gamma_n},
		\label{decay_width}
}
where $\Gamma_n$ is the decay rate of the $\varphi_n$ resonance.
Under a mild assumption that the decay rate increases with $m_n$ at least linearly, effect of such a decay width can be taken into account by slightly shifting the contour of the large $s$ limit: $s\to(1+i\epsilon)\infty$ where the positive constant $\epsilon$ can be taken arbitrary small but must be kept finite.\footnote{
We note that in our model, the decay rate of the resonance into $W^\pm$ pair is indeed sizable already at the lowest KK Higgs mode~\cite{Haba:2009pb}:
\als{
\Gamma_{\varphi_1\to W^+W^-}
	&=	\paren{2\sqrt{2}\over\pi}^2{g_4^2m_H^3\over64\pi m_W^2}\paren{1-{2m_W^2\over m_H^2}}^2\sqrt{1-{4m_W^2\over m_H^2}},
}
where we note that the mass of this first KK Higgs, the ``Higgs impostor,'' is exactly the KK scale: $m_H=1/R$.
}
This type of limit is taken when we get the Regge and hard scattering limits from the tree-level string amplitude. See e.g.~\cite{Matsuo:2008fj} for more detailed discussion. 
By this prescription, we get the exponential limit:
\al{
\tan{\pi R\sqrt{s}\over2}
	\to	-1,
}
and finally
\al{
\mathcal{M}_{W^+_LW^-_L\to W^+_LW^-_L}^\text{KK Higgs exchange}
	&\to
		-{s\paren{1+\cos\theta}\over2\vEW^2}
		-{\sqrt{2s}\over\vEW^2\pi R}\paren{\sqrt{2}+\sqrt{1-\cos\theta}}
		+\mathcal{O}(s^0).
}
That is, the total amplitude becomes
\al{
\mathcal{M}_{W^+_LW^-_L\to W^+_LW^-_L}
	&\to
		-{\sqrt{2s}\over\vEW^2\pi R}\paren{\sqrt{2}+\sqrt{1-\cos\theta}}
		+\mathcal{O}(s^0).
		\label{total_amplitude}
}

A differential cross section in CM frame is written, when all the masses for incoming and outgoing four particles are equal, as
\al{
{d\sigma\over d\Omega}
	&=	{1\over64\pi^2s}\ab{\mathcal{M}}^2,
}
and we get the elastic cross section that takes the dominant KK Higgs-exchange contribution into account:
\al{
\sigma_{W^+_LW^-_L\to W^+_LW^-_L}
	=	2\pi\int_{-1}^1d\cos\theta{d\sigma\over d\Omega}
	=	{1\over32\pi s}\int_{-1}^1d\cos\theta\ab{\mathcal{M}}^2
	\to	{17\over24\pi^3\vEW^4R^2}.
}
We see that the tree-level elastic cross section remains constant in the high energy limit and hence is marginally unitarized.

\cred{
In the literature the question of the unitarity of $WW$ scattering is typically addressed using the Nambu--Goldstone (NG) boson equivalence theorem. Following~\cite{Haba:2009pb,Haba:2010xz}, one may speculate that the NG boson that is absorbed by the gauge zero mode $W^\mu_0$ is an infinite sum: $\chi_\text{NG}=\sum_\text{$n$: odd}{2\over n\pi}\wt\chi_n$, with each $\wt\chi_n$ being a linear combination of $\chi_n$ and $W^z_n$. To prove that, one has to compute an infinite number of KK-number violating scattering amplitudes and sum them up correctly. In this paper we have restricted ourselves to the simpler analysis as is presented above.
}

We have found the growing amplitude with energy $\mathcal M\propto \sqrt{s}$ after summing over infinite KK modes, though the original amplitude is expanded as Eq.~\eqref{general_expansion} and does not have such half power of $s$, where $\mathcal{M}^{(4)}$ cancels within SM gauge amplitudes, while we have seen that $\mathcal{M}^{(2)}$ cancels between the sum of SM gauge amplitudes~\eqref{SM_gauge_only} and that of the KK Higgs amplitudes~\eqref{summed_amplitude_in_limit}. This half power arises when one sums over infinite KK modes and can be interpreted as follows.\footnote{\cred{
The half power $\sqrt{s}$ resides within the terms proportional to (hyperbolic) tangent in Eq.~\eqref{exact_sum}. The poles $\sqrt{s}=1/R, 3/R, \dots$ in tangent correspond to the resonances as is explained in Eq.~\eqref{decay_width}. In KK picture, the half power could be interpreted as the effect of taking into account the width. This behavior should appear even when one considers scattering with Euclidean external momenta.
}}
When we sum over infinite KK modes, we see a scattering within full five-dimensional bulk. In five dimensions, the gauge coupling has mass dimension $[g]=-1/2$ and hence from naive dimension counting, we expect
\al{
\mathcal M^\text{naive}
	\sim
		g^2\sqrt{s}
%	=	{g_4^2L\sqrt{s}}
%	=
%		{4m_W^2L\sqrt{s}\over\vEW^2}
	\sim
		{\sqrt{s}\over\vEW^2R}{m_W^2\over\mKK^2},
}
which is what we have found in Eq.~\eqref{total_amplitude}, up to the extra factor $\mKK^2/m_W^2$ to be multiplied.

%%%%%%%%%%%
\subsection{Partial-wave unitarity}
%%%%%%%%%%%
Let us expand the $W^+_LW^-_L$ scattering amplitude into partial-waves
\al{
\mathcal{M}(s,\cos\theta)
	&=	\sum_{J=0}^\infty\paren{2J+1}\mathcal{M}_J(s)P_J(\cos\theta),
}
where the $J$th partial amplitude is obtained inversely
\al{
\mathcal{M}_J(s)
	&=	{1\over2}\int_{-1}^1d\cos\theta\,P_J(\cos\theta)
		\mathcal{M}(s,\cos\theta).
		\label{partial_wave_amplitude}
}
In the high energy limit~\eqref{total_amplitude}, we get
\al{
\mathcal{M}_J
	&=	-{\sqrt{s}\over\vEW^2\pi R}c_J,
}
where
\al{
c_J	&=	\int_{-1}^1d\cos\theta\,P_J(\cos\theta)\paren{
			1+\sqrt{1-\cos\theta\over2}
			}.
}
Concretely,  $c_J={10\over3},-{4\over15},-{4\over105}\dots$ for $J=0,1,2, \dots$, respectively.

The unitarity of the partial-wave amplitude reads
\al{
\im{\mathcal{M}_J}
	&\geq	{\ab{\bs k}\over 8\pi\sqrt{s}}\ab{\mathcal{M}_J}^2
	\to	{1\over16\pi}\ab{\mathcal{M}_J}^2,
		\label{partial_wave_unitarity_exact}
}
where high energy limit $s\gg m_W^2$ is taken in the last step.
At the tree level, we do not have the imaginary part at all and it is customary to use a corollary of the exact unitarity condition~\eqref{partial_wave_unitarity_exact}:
\al{
1	&\geq	{\ab{\bs k}\over 8\pi\sqrt{s}}\ab{\mathcal M_J}
	\to	{1\over16\pi}\ab{\mathcal{M}_J}.
}
This way, the tree-level partial-wave unitarity condition is, for the most stringent $J=0$ partial-wave amplitude,
\al{
\sqrt{s}<{24\pi^2\vEW^2\over5\mKK}=:\Lambda,
}
where $\mKK:=1/R$ is the first KK Higgs mass. Around the scale $\Lambda$, higher loop corrections become important in the scattering, though the gauge theory itself is still well defined as we can show that our theory possesses a nilpotent BRST symmetry. When we require that there exists a weak coupling region for, say, three KK modes: $\Lambda\gtrsim3\mKK$, we get
\al{
\mKK\lesssim980\,\text{GeV}.
}
More concretely, for KK scales favored by the electroweak precision data within 90\% CL~\cite{Haba:2009pb}: $\mKK=430$--$500\,\text{GeV}$, we get
\al{
\Lambda=6.7\text{--}5.7\,\text{TeV},
	\label{our_cutoff}
}
which are well beyond the corresponding KK scales, at least ten KK modes being within tree-level unitarity range.

\cred{In this paper, we have concentrated on the elastic channels. In an analysis of a Higgsless model~\cite{Papucci:2004ip}, inclusion of the inelastic channels into KK $W$ bosons leads to a lower 5D cutoff:
\al{
\Lambda_5\sim \Lambda_4\sqrt{N_\text{KK}}
	\label{papucci_bound}
}
than considering only the elastic ones in the Higgsless model, where $N_\text{KK}$ is the number of KK modes below the 5D cutoff and $\Lambda_4\simeq2\,\text{TeV}$ is the cutoff of the four dimensional SM without Higgs. (The relation~\eqref{papucci_bound} is consistent with the 5D Naive Dimensional Analysis.) In Higgsless models, the second KK states must be much heavier than twice the first KK mass to match the electroweak constraint. In contrast, our model has the equal separation of the KK modes without contradicting to the electroweak data: $N_\text{KK}=\Lambda_5/m_\text{KK}$. Therefore, the relation~\eqref{papucci_bound} simply leads to $\Lambda_5\sim 8\,\text{TeV}$ for $m_\text{KK}\simeq 500\,\text{GeV}$ in our case, which is the same order as Eq.~\eqref{our_cutoff}.}

%%%%%%%%%%%%%%%%%
\section{Summary}
%%%%%%%%%%%%%%%%%
We have briefly sketched how the five-dimensional UED model, compactified on a line segment, is consistently formulated when the EWSB is solely due to the non-zero Dirichlet boundary conditions on the bulk Higgs field, in the limit of vanishing bulk and boundary potentials. We have discussed how the elastic scattering of the longitudinal $W^+W^-$ zero modes is unitarized, under the absence of the Higgs zero mode, by showing that the sum over the contribution of infinite tower of the KK Higgs modes exactly cancels the $\mathcal{O}(s)$ contribution from the SM gauge sector. Further, we have obtained the high energy limit taken \emph{after} summing over all the KK Higgs modes, that exhibit the behavior $\mathcal{M}\propto \sqrt{s}$, which never appears in four-dimensional level before summation and is genuinely five-dimensional. Resultant tree-level partial-wave unitarity condition leads, for a range favored by the electroweak precision data within 90\% CL $\mKK=430$--$500\,\text{GeV}$, to the strongly-coupled UV-cutoff scale $\Lambda=6.7$--$5.7\,\text{TeV}$, which is well above the KK scale. Details of our study and further discussions will be presented in a separate publication~\cite{NishiwakiOda}.

%%%%%%%%
\subsection*{Acknowledgment}
%%%%%%%%
We are most grateful to Alex Pomarol for valuable comments. \cred{We appreciate earlier discussions with Naoyuki Haba that brought attention to the unitarity issues on the model. We also thank Tomohiro Abe,} Arthur Hebecker, Victor Kim, C.S.\ Lim, Hitoshi Murayama, Makoto Sakamoto, Marco Serone, and Ryo Takahashi for useful discussions and Gianmassimo Tasinato, Yasuhiro Yamamoto and Ivonne Zavala for helpful conversations. K.O.\ acknowledges the hospitality of the particle theory group of Bonn University while this work is partly developed. The stay of K.O.\ in Bonn University and CERN is financially supported in part by the JSPS International Training Program of Osaka University. K.O.\ is partially supported by Scientific Grant by Ministry of Education and Science (Japan), Nos.~19740171 and 20244028.

\end{fmffile}

\bibliographystyle{TitleAndArxiv}
%\bibliography{paper}

\providecommand{\bysame}{\leavevmode\hbox to3em{\hrulefill}\thinspace}

\end{document}